\documentclass[presentation,8pt]{beamer}
\usepackage[T2A]{fontenc}%
\usepackage[cp1251]{inputenc}
\usepackage[russian,english]{babel}%
\newcommand{\cat}[1]{\left|#1\right\rangle}
\newcommand{\ecat}[1]{\left|#1\right)}
\newcommand{\catt}[1]{\left|\left. #1 \right\rangle\right\rangle}
\newcommand{\ero}[2]{\left|#1\right)\left(#2\right|}
\newcommand{\pro}[2]{\left|\left.#1\right\rangle\right\rangle \left\langle\left\langle#2\right.\right|}
\newcommand{\ro}[2]{\left|#1\right\rangle\left\langle#2\right|}
\newcommand{\roa}[1]{\left|#1\right\rangle\left\langle#1\right|}
\newcommand{\eroa}[1]{\left|#1\right)\left(#1\right|}
\newcommand{\aver}[1]{\left\langle #1\right\rangle}
\newcommand{\mess}[2]{\left\langle #1\left|#2\right|#1\right\rangle}
\newcommand{\Tr}[1]{Tr\left(#1\right)}
\newcommand{\proa}[1]{\left|\left.#1\right\rangle\right\rangle \left\langle\left\langle#1\right.\right|}

\title{Complexity of measurement of a qubit pair}

\author{Constantin V. Usenko}%
\institute[Kyiv]{Kyiv Schevchenko university}%

\date{}

\begin{document}

\begin{frame}

\maketitle

 \tableofcontents
\end{frame}

\section{Introduction}
\subsection{An Implementation of Qubit pair}
%FRAME 2
\begin{frame}{Two-level atom and mode of quantum field}
\begin{itemize}
	\item Atom:\hfill $\cat{0,1}$,\hfill $\sigma_3$, $\sigma_\pm$ \hfill \hspace*{2em}
	\item Field:\hfill $\left\{\ecat{n};\ n=0,\ldots,\infty\right\}$,\hfill $\hat{a}$,  $\hat{a}^+$,  $\hat{n}=\hat{a}^+\hat{a}$  \hfill \hspace*{2em}
	\item Composite system: $\left\{\catt{k,n}=\ecat{n}\otimes\cat{k};\ k=0,1;n=0,\ldots,\infty\right\}$
\end{itemize}
Dynamics
\[\hat{H}=\frac{\hbar\omega}{2}\sigma_3+\hbar\omega\hat{a}^+\hat{a}+i\hbar\xi\left(\hat{a}^+\sigma_+-\hat{a}\sigma_-\right) \]
\[\catt{\phi\left(t\right)}
=\sum_{n=0}^\infty{\ecat{n}\otimes\left(\phi_{0,n}\left(t\right)\cat{0}+\phi_{1,n}\left(t\right)\cat{1}\right)}
\]
\[\catt{\phi\left(t\right)}
=\cos\theta\left(t\right)\ecat{\psi_0\left(t\right)}\otimes\cat{0}+\sin\theta\left(t\right)\ecat{\psi_1\left(t\right)}\otimes\cat{1}
\]
{\tiny
\[\phi_{0,n}\left(t\right)=\cos\theta\left(t\right)\psi_{0,n}\left(t\right)=\phi_{init:0,n}\cos\omega_{n-1}t-i\phi_{init:1,{n-1}}\sin\omega_{n-1}t \]
\[\phi_{1,n}\left(t\right)=\sin\theta\left(t\right)\psi_{1,n}\left(t\right)=-i\phi_{init:0,n+1}\sin\omega_nt+\phi_{init:1,n}\cos\omega_nt \]
}
\begin{itemize}
	\item Second qubit: \hfill $\ecat{\psi_0\left(t\right)},\ecat{\psi_1\left(t\right)}$,\hfill  $\hat{s}_+=\ero{\psi_1\left(t\right)}{\psi_0\left(t\right)}$ \hfill \hspace*{2em}
\end{itemize}
A state of field, like a state of a qubit, in the case of exactly determined instant of termination of field-atom interaction, belongs to two-dimensional space.  %Состояние поля, подобно состоянию кубита, в случае точно определенного момента прекращения взаимодействия поля с атомом принадлежит двумерному пространству.
\end{frame}%FRAME 2

\subsection{Complexity of Measurement}
%FRAME 3
\begin{frame}{Definition and Properties}
\begin{flushright}
 {\footnotesize %C. V. Usenko 
 2013 {\it Journal of Physics: Conference Series} {\bf V 442} 012061}
\end{flushright}

Complexity of Measurement is characterized by the number of events of measurements 
 %Измерительная сложность характеризуется количеством актов измерений
  $M\left(\epsilon\right)$, enough to come to given accuracy %достаточных для получения заданной точности
   $\sigma_X\leq \epsilon$:
\begin{equation}
	C_M\left(\epsilon\right)=\log_2 M\left(\epsilon\right).
\end{equation}

Joint measurement 
%add
or correlated parts of system
\[C_M\left(O_1 \& O_2;\epsilon\right)={C_M\left(O_1 ;\epsilon\right)}+{C_M\left( O_2;\epsilon\right)}
\]
Successive  measurement
%add
or incompatible observables
\[C_M\left(O_1 , O_2;\epsilon\right)=\log_2\left(2^{C_M\left(O_1 ;\epsilon\right)}+2^{C_M\left( O_2;\epsilon\right)}\right)
\]
%\end{frame}
Measurement of N 
%add
incompatible 
%add
observables with equal complexities
\[\forall_n C_M\left(O_n ;\epsilon\right)=C_M\left(O ;\epsilon\right)\quad\mapsto\quad C_M\left(O_1 ,\ldots, O_N;\epsilon\right)=\log_2 N+C_M\left(O ;\epsilon\right)
\]

\end{frame}%FRAME 3

%FRAME 4
\begin{frame}{Determination }
\begin{block}{$\dim \mathcal{H}=2$}
Practical formula for evaluation of complexity of measurement of probability %Практическая формула для оценки сложности измерения вероятности
 p with error 
 %с погрешностью
  s:
%Complexity of Measurement
\begin{equation}
	C_p\left(p ,s\right)=\log_2\left(p\left(1-p\right)\right)-2\log_2 s.\label{Cps}
\end{equation}
The first term of this sum characterizes contribution from the measured value. Maximal value of this term is achieved at %Первый член этой суммы характеризует вклад измеряемого значения. Максимальное  значение этого члена достигается при
 $p=1/2$ and exactly compensates the second term in the case of maximal possible error % и в точности компенсирует второй член в случае максимальной возможной погрешности
  $s=1/2$. 
\end{block}
\begin{block}{$\dim \mathcal{H}=N$}
Practical formula for evaluation of the complexity of measurement of % Практическая формула для оценки сложности измерения
 N-dimensional probability distribution with error % distribution мерного распределения вероятностей с погрешностью
  s:

\begin{equation}
	-\log_2 N-2\log_2 s \leq  C_N\left(s\right) \leq -4-2\log_2 s.\label{Cns}
\end{equation}
The infimum corresponds to uniform probability distribution %Нижняя граница соответствует равномерному распределению вероятностей
 ($p=1/N$),  the supremum corresponds to the maximal value of complexity of series of measurements %верхняя граница соответствует максимальному значению сложности  серии измерений
  ($p_{max}=1/2$).
\end{block}

\end{frame}%FRAME 4

%FRAME 5
\subsection{Measurement of a State}
\begin{frame}{Description of State}
 Basis
\[\begin{array}{l}
\catt{1}
=\cos\theta\ecat{0}\otimes\cat{0}-\sin\theta\ecat{1}\otimes\cat{1}\\
\catt{2}
=\sin\theta\ecat{0}\otimes\cat{0}+\cos\theta\ecat{1}\otimes\cat{1}\\
\catt{3}
=\cos\vartheta\ecat{1}\otimes\cat{0}-\sin\vartheta\ecat{0}\otimes\cat{1}\\
\catt{4}
=\sin\vartheta\ecat{1}\otimes\cat{0}+\cos\vartheta\ecat{0}\otimes\cat{1}\\
\end{array}
\]
 Density matrix
\[\hat{\rho}=\sum_{k,k'=1}^4\rho_{k,k'}\pro{k'}{k}=\sum_{a,a',f,f'=1}^2\rho_{a,f;a',f'}\ro{a'}{a}\otimes\ero{b'}{b}
\]
\[\hat{\rho}=\left(
\begin{array}{cccc}
p_1&\rho_{1,2}&\rho_{1,3}&\rho_{1,4}\\
\rho^*_{1,2}&p_2&\rho_{2,3}&\rho_{2,4}\\
\rho^*_{1,3}&\rho^*_{2,3}&p_3&\rho_{3,4}\\
\rho^*_{1,4}&\rho^*_{2,4}&\rho^*_{3,4}&p_4
\end{array}
\right)\qquad p_1+p_2+p_3+p_4=1
\]
%Измерение состояния осуществляется посредством определения всех 16 компонент матрицы плотности. Независимых компонент 9. 
By the results of measurements it is needed to determine: %По результатам измерений необходимо определить:

\begin{itemize}
	\item If the eigen-basis of the system is known, 3 independent values; % Если известен собственный базис системы, 3 независимых величины;
	\item If the eigen-bases of the sub-systems are known, 5 independent values; %Если известны собственные базисы подсистем, 5 независимых величин;
	\item In general case, 9 independent values % В общем случае 9 независимых величин.
\end{itemize}
\end{frame}%FRAME 5

%FRAME 6
\begin{frame}{System and Local Observables }

Measurement of state of a qubit pair can be realized by means of measurement of local observables only. % Измерение состояния пары кубитов может осуществляться только посредством измерения локальных наблюдаемых.

\begin{itemize}
	\item 
Atom observables 
\[ \sigma_1=\ro{1}{0}+\ro{0}{1},\quad \sigma_2=-i\ro{1}{0}+i\ro{0}{1}
\]
\[\sigma_3=-\roa{0}+\roa{1},\quad \hat{I}_\sigma =\roa{0}+\roa{1}
\]
	\item 
Mode observables 
{%\small
\[ s_1=\ero{\psi_\bot}{\psi_0}+\ero{\psi_0}{\psi_\bot},\quad  s_2=-i\ero{\psi_\bot}{\psi_0}+i\ero{\psi_0}{\psi_\bot}
\]
\[s_3=-\eroa{\psi_0}+\eroa{\psi_\bot},\quad   \hat{I}_s=\eroa{\psi_0}+\eroa{\psi_\bot}
\]
}
	\item 
Observables of composite system
\[O_{system=}\mathcal{L}\left(O_\sigma\otimes O_s\right)\]
\[\begin{array}{cccc}
\hat{I}_\sigma\otimes \hat{I}_s &\hat{I}_\sigma\otimes s_1 & \hat{I}_\sigma\otimes s_2 & \hat{I}_\sigma\otimes s_3 \\ 
\sigma_1\otimes \hat{I}_s &\sigma_1\otimes s_1 & \sigma_1\otimes s_2 & \sigma_1\otimes s_3 \\ 
\sigma_2\otimes \hat{I}_s &\sigma_2\otimes s_1 & \sigma_2\otimes s_2 & \sigma_2\otimes s_3 \\ 
\sigma_3\otimes \hat{I}_s &\sigma_3\otimes s_1 & \sigma_3\otimes s_2 & \sigma_3\otimes s_3 \\ 
\end{array}
\]
\end{itemize}
\end{frame}%FRAME 6

%FRAME 7
\begin{frame}{Measured values}
Eigen-basis and Eigenvalues
\[\hat{O}\cat{k}=O_k\cat{k}\qquad \left|\hat{O}\mapsto\quad \mathcal{F}\left(\hat{O}\right)\exists \mathcal{F}:\quad  \mathcal{F}\left(O_k\right)=k\right| \qquad \hat{O} =\sum_{ k=1}^N k\roa{k}
\]
Measurement of all the moments of an observable is equivalent to the measurement of probability distribution % Измерение всех моментов наблюдаемой эквивалентно измерению распределения вероятностей
\[\hat{P}_k=\roa{k};\qquad \left\{\aver{\hat{O}^n}=\sum_{k=1}^NO_k^np_k,n=1\ldots N-1\right\} \equiv \left\{\aver{\hat{P}_k}=p_k,k=1\ldots N-1\right\}
\]
%Физически набор $\hat{P}_k$ является набором взаимно дополнительных  совместных  счетчиков

Mathematical model of a 
%delete 
%measuring instrument
%insert
detector
 - a non-degenerate observable, or an equivalent set of counters: %Математическая модель измерительного прибора -- невырожденная наблюдаемая  или эквивалентный набор счетчиков

\[\hat{O} =\sum_{ k=1}^N k\roa{k} \quad \equiv \quad \left\{\hat{P}_k=\roa{k},k=1,\ldots,N\right\}
\]
Result of measurement - a set of probabilities % Результат измерения -- набор вероятностей
 $\left\{p_k=\mess{k}{\hat{\rho}},\ k=1,\ldots,N\right\}$, equal to diagonal elements of density matrix % равных диагональным элементам матрицы плотности
  $p_k=\rho_{k,k}$.
\end{frame}%FRAME 7

%FRAME 8
\begin{frame}{Incompatibility}

Non-diagonal components of density matrix can not be measured.

\(\left[\sigma_x\sigma_y\right]\neq 0\) \hfill Non-commuting observables correspond to different
%delete 
%instruments,
%insert
detectors,
 as a result of measurement there is a set of elements of density matrix diagonal in some other basis. Measurement with a large enough set of 
% instruments 
detectors
 produces a set of results enough for calculation of all the components of density matrix. 
 
 % Некоммутирующие наблюдаемые соответствуют разным приборам, результатом измерения является набор элементов матрицы плотности, диагональных в другом базисе. Измерение достаточно большим набором приборов продуцирует множество реультатов, достаточное для вычисления всех компонент матрицы плотности.

 \begin{block}{Minimal Set of Observables}

The minimal set consists of $N+1$ observables. i.e. 3 observables for qubit, 5 observables for a qubit pair.

Let $N_a$ is dimension of the first subsystem and $N_f$ -- of the second one. The number of possible pairs of needed observables % Количество возможных пар необходимых наблюдаемых
  $\left(N_a+1\right)\left(N_f+1\right)$ exceeds the needed number of observables of a composite system %превышает необходимое количество наблюдаемых композитной системы
   $N_aN_f+1$.

$\left(N_a+N_f\right)$ pairs of local observables can be excluded from the process of measurement. % пар локальных наблюдаемых можно исключить из процесса измерения.  

 In the case of a pair of qubits measurement of only 5 of 9 pairs of observables is enough. % В случае пары кубитов можно  измерять только 5 из 9 пар наблюдаемых.

\begin{flushright}
 {\footnotesize %C. V. Usenko 
 2011 {\it Optics and Spectroscopy} {\bf 111}  678 }
\end{flushright}
\end{block}
 \begin{alertblock}{What pairs of local observables for the first and the second qubits are to be measured?} %Какие пары локальных наблюдаемых первого и второго кубитов нужно измерять?}
\end{alertblock}
\end{frame}%FRAME 8

%FRAME 9
\begin{frame}{Measurement of Qubit States}
Qubit: \(\hat{O} =O_0\roa{0}+O_1\roa{1}\quad \equiv \quad \left\{\hat{P}_0=\roa{0},\hat{P}_1=\roa{1}\right\}\).
Measurement of each of observables determines one value only. %Измерение каждой наблюдаемой определяет только одну величину. 

An arbitrary density matrix is given by a linear combination of Pauli matrices  % Произвольная матрица плотности представляется линейной комбинацией матриц Паули

\[\hat{\rho}=\frac{1}{2}\hat{I}+\frac{\Delta p}{2}\hat{\sigma}_3 +\frac{d'}{2}\hat{\sigma}_1+\frac{d''}{2}\hat{\sigma}_2\]
The problem on reconstruction of a state is solved by the results of a series of measurements for 3 incompatible observables %Задача восстановления состояния решается по результатам серий измерений 3 несовместимых наблюдаемых 
\[\Delta p =\Tr{\hat{\sigma}_3 \hat{\rho}},\quad d' =\Tr{\hat{\sigma}_1 \hat{\rho}},\quad d''=\Tr{\hat{\sigma}_2 \hat{\rho}} \]
Complexity of Measurement 

\[ C_2\left(\epsilon\right)=\log_2\left(3-\Delta p^{2}-d'^2-d''^2\right)-2\log_2 s
\]
\[ 1-2\log_2 s \leq C_2\left(\epsilon\right)\leq\log_2 3-2\log_2 s
\]
The infimum corresponds to pure state, the supremum corresponds to uniform probability distribution. % Нижняя граница соответствует чистому состоянию,  верхняя граница соответствует равномерному распределению вероятностей.

\end{frame}%FRAME 9

%FRAME 10
\begin{frame}{Measurement of Ququart States}

Pair of Qubits:

A non-degenerate observable
 \(\hat{O} =\displaystyle\sum_{k=1}^4O_k\roa{k}\quad \equiv \quad \left\{\hat{P}_k=\roa{k},k=1,2,3,4\right\}\). Measurement produces   3 independent values 
$p_1+p_2+p_3+p_4=1$

A density matrix contains 15 independent values, so measurement of 5 non-degenerate observables is needed. 

Typically 
\[\hat{J}_+=\sqrt{3}\ro{2}{1}+2\ro{3}{2}+3\ro{4}{3},\quad  
\hat{J}_-=\sqrt{3}\ro{1}{2}+2\ro{2}{3}+\sqrt{3}\ro{3}{4}\]
\[\hat{J}_0=-\frac{3}{2}\roa{1}-\frac{1}{2}\roa{2}+\frac{1}{2}\roa{3}+\frac{3}{2}\roa{4}\]
\[\hat{J}_k=e^{\frac{-i\pi}{2} \left(k-1\right)}\hat{J}_++e^{\frac{i\pi}{2} \left(k-1\right)}\hat{J}_-,\quad k=1,2,3,4
\]

%Complexity of Measurement 

%\[ C_M\left(\epsilon\right)=\log_2\left(3-\Delta p^{2}-d'^2-d''^2\right)-2\log_2 s.\label{Cps}
%\]

Complexity of Measurement: 

\[ 
	\log_2 \frac{5}{4}+\log_2 \frac{3}{4}-2\log_2 s \leq  C_4\left(s\right) \leq \log_2 \frac{5}{4}-2\log_2 s.
\]
Error that does not exceed 0.1 can be obtained in 5 series of measurements, from 19 to 25 events of measurement each. %  Погрешность, не превышающая 0.1, может быть получена при проведении 5 серий измерений  длиной от 19 до 25 событий измерения каждая. 

\end{frame}%FRAME 10

%FRAME 11
\begin{frame}{Composite states}

$\cat{\psi}$ represents the subsystem A(tom), $\ecat{\alpha}$ represents the subsystem F(ield), $\catt{\Psi}$ represents the S(ystem).

State of System, representation in the basis of eigenvectors: % представление в базисе собственных векторов:
\begin{equation}
	\hat{\rho}=\sum_{n=1}^4p_m\proa{m}
\end{equation}
Schmidt decompositions of eigenvectors of density matrix: 

\begin{equation}\label{parabasis}
	\begin{array}{l}
	\catt{1}\equiv\catt{Cat:0}=\cos\theta_c\cat{0}\otimes\ecat{0}+\sin\theta_c\cat{1}\otimes\ecat{1},\\
	\catt{2}\equiv\catt{Cat:1}=-\sin\theta_c\cat{0}\otimes\ecat{0}+\cos\theta_c\cat{1}\otimes\ecat{1};\\
	\catt{3}\equiv\catt{EPR:0}=\cos\theta_e\cat{0}\otimes\ecat{1}+\sin\theta_e\cat{1}\otimes\ecat{0},\\
	\catt{4}\equiv\catt{EPR:1}=-\sin\theta_e\cat{0}\otimes\ecat{1}+\cos\theta_e\cat{1}\otimes\ecat{0}.
	\end{array}
\end{equation}
Phase factors of all the coefficients are removed by special selection of phase factors of the basis vectors of subsystems. % Фазовые множители всех коэффициентов устранены специальным выбором фазовых множителей базисных векторов подсистем

Density matrices of subsystems are diagonal:

\begin{equation}
	\begin{array}{l}
	\hat{\rho}_a=p_{a:0}\roa{0}+p_{a:1}\roa{1},\\
	\hat{\rho}_f=p_{f:0}\eroa{0}+p_{f:1}\eroa{1},
	\end{array}
\end{equation}

\end{frame}%FRAME 11

%FRAME 12
\begin{frame}{Joint Probabilities of Local Observables:}

\begin{block}{Solvable system of equations for parameters of density matrix  $p_1$, $p_2$, $p_3$, $p_4$,  $\theta_c$, $\theta_e$.
} 
Only the values of probabilities with non-trivial correlation are given :% Приведены только значения вероятностей с нетривиальной корреляцией 

$\sigma_3 \ \& \ \hat{s}_3$
\begin{equation}
	\begin{array}{ll}
	p^{\left\{33\right\}}_{0,0}=\frac{p_1+p_2}{2}+\frac{p_1-p_2}{2}\cos2\theta_c,&
	p^{\left\{33\right\}}_{1,1}=\frac{p_1+p_2}{2}-\frac{p_1-p_2}{2}\cos2\theta_c,\\
	p^{\left\{33\right\}}_{0,1}=\frac{p_3+p_4}{2}+\frac{p_3-p_4}{2}\cos2\theta_e,&
	p^{\left\{33\right\}}_{1,0}=\frac{p_3+p_4}{2}-\frac{p_3-p_4}{2}\cos2\theta_e.
	\end{array}
\end{equation}

 ${\sigma}_{1} \ \& \ \hat{s}_{1}$ 
\begin{equation}
	\begin{array}{l}
	p^{\left\{11\right\}}_{00}=p^{\left\{11\right\}}_{11}=
	\frac{1}{4}+\frac{p_1-p_2}{4}\sin2\theta_c+\frac{p_3-p_4}{4}\sin2\theta_e,\\
		p^{\left\{11\right\}}_{0,1}=p^{\left\{11\right\}}_{1,0}=
		\frac{1}{4}-\frac{p_1-p_2}{4}\sin2\theta_c-\frac{p_3-p_4}{4}\sin2\theta_e,
		\end{array}
\end{equation}
 ${\sigma}_{2} \ \& \ \hat{s}_{2}$ 
\begin{equation}
	\begin{array}{l}
p^{\left\{22\right\}}_{00}=p^{\left\{22\right\}}_{11}=
\frac{1}{4}-\frac{p_1-p_2}{4}\sin2\theta_c+\frac{p_3-p_4}{4}\sin2\theta_e,\\
		p^{\left\{22\right\}}_{01}=p^{\left\{22\right\}}_{10}=
		\frac{1}{4}+\frac{p_1-p_2}{4}\sin2\theta_c-\frac{p_3-p_4}{4}\sin2\theta_e, \\
		\end{array}
\end{equation}
%Разрешимая система (solvable system) уравнений относительно параметров матрицы плотности

\end{block}
Previous knowledge on optimal set of local bases makes it possible to decrease the number of the pairs of observables to be measured to 3. % Предварительное знание оптимального набора локальных базисов позволяет уменьшить  количество подлежащих измерению пар наблюдаемых до 3. 

Complexity of measurement: 
\[ 
	  C_4\left(s\right) = \log_2 3 -2-2\log_2 s.
\]
Error that does not exceed 0.1 can be obtained in 3 series of measurements, 25 events of measurement each. % Погрешность, не превышающая 0.1, может быть получена при проведении 3 серий измерений  по 25 событий измерения каждая. 

\end{frame}%FRAME 12

\section{Process of Measurement}
\subsection{Events and Sequences}

%FRAME 13
\begin{frame}{Sequence of Events of Measurement}

%\note
{At measurement of an observable statistical character of observables of quantum physics makes need in large enough series of events of measurement. % Статистический характер наблюдаемых квантовой физики делает необходимым при измерении любой наблюдаемой осуществление достаточно большой серии событий измерения.
 }

\begin{block}{Preparator}

performs selection of a pure state $\cat{\psi_n} \in \left[\cat{\psi_1},\ldots.\cat{\psi_N}\right]$. 

Result: %Результат:   
\[\mathcal{S}=\left\{\cat{\psi_{n_1}},\ldots,\cat{\psi_{n_k}},\ldots,\cat{\psi_{n_K}}\right\}\equiv 
S=\left\{n_1,\ldots,n_k,\ldots,n_K\right\}\]

\end{block}
\note{In each event the measured system is prepared by the 
%instrument -- source 
Source
in one state from the set of possible states
%В каждом событии измеряемая система приготавливается прибором -- источником в одном состоянии из набора возможных состояний
}

\note{The result of preparation is a sequence of states
%Результатом приготовления является последовательность состояний
}
\begin{block}{Registrator}
 performs selection of the 
 %delete
 %measuring instrument
%insert
detector
 % осуществляет выбор измеряющего прибора 
 $\hat{O} \in \left\{\hat{O}^{\left\{d\right\}},d=0,\ldots,N\right\}$. 
 
The result is in a sequence of pairs 
%Результат - последовательность пар 
[%instrument
detector number 
%номер детектора 
$d$,  measured value %измеренное значение 
$m$]:
\[R=\left\{\left[d_1,m_1\right],\ldots,\left[d_k,m_k\right],\ldots,\left[d_K,m_K\right]\right\}\]

The result of measurements with a specific 
%instrument
detector
 number %Результат измерений  конкретным прибором номер 
$d_s$ -- a subsequence in which only the events of measurement with that
% instrument 
detector
are present. %подпоследовательность, в которой присутствуют только события измерения данным прибором. 
\[R_s=\left\{\left[d_k=d_s,m_k\right]\right\}\]

\end{block}

\end{frame}%FRAME 13

%FRAME 14
\begin{frame}{Types of Processes of Measurement I}
Sequence of prepared states produces the density matrix of Source $\hat{\rho}_S$  
\[\hat{\rho}_S=\frac{1}{K}\sum_{k=1}^K\roa{n_k}=\sum_{n=1}^{N}\frac{K_{n}}{K}\roa{n} \]
\begin{block}{Analysis}
Preparator and Registrator realize  independent series of measurement for each state and each device in coordination. %  согласованно осуществляют независимые серии измерений для каждого состояния и каждого прибора. 
Sequences of states measured by d-th detector  produce density matrices of measured states $\hat{\rho}_M^{\left\{d\right\}}$
\end{block}
\[ \left\{\hat{\rho}_M^{\left\{d\right\}}=\frac{1}{K_d}\sum_{k=1}^{K_d}\roa{n_{k|d}} =\sum_{n=1}^{N}\frac{K_{n|d}}{K_d}\roa{n} 
,d=0,\ldots,N\right\}\]

Those density matrices are same only in the case if Preparator provides same repetition frequencies 
%Эти матрицы плотности одинаковы только в том случае, если Препаратор обеспечивает одинаковые частоты заполнения 
$\frac{K_{n|d}}{K_d}$ for each series. %в каждой серии.
\end{frame}%FRAME 14

%FRAME 15
\begin{frame}{Types of Processes of Measurement II}
\begin{block}{Data transfer}
Registrator uses a %device
detector
that performs non-demolition measurements. The sequence of results exactly repeats the initial
% использует один прибор, осуществляющий неразрушающие измерения. Последовательность результатов точно повторяет исходную
  $\mathcal{R}=\mathcal{S}$. The density matrix is characterized by repetition frequencies.

  % Матрица плотности характеризуется частотами повторений. Одной матрице плотности

\[\hat{\rho}=\sum_{n=1}^N\frac{K_n}{K}\roa{n}\quad\mapsto\quad M_I=\frac{K!}{K_1!\ldots K_N!}\]
 One density matrix corresponds to $M_I$ different sequences. % разных последовательностей.
 
\begin{flushright}
 { Control of accuracy is performed by outside methods. }% Контроль точности осуществляется сторонними методами.
\end{flushright}
\end{block}
\begin{block}{Eavesdropping}
Difference between the density matrices of measured states %Различие между матрицами плотности измеренных состояний 
$\hat{\rho}_M^{\left\{d\right\}}$ and the density matrix of the Source % и  матрицей плотности Источника 
$\hat{\rho}_S$ can not be determined without coordination of choice between Preparator and Registrator. %не может буть определено без согласования выборов между Preparator and Registrator. 

Assumption that one of the 
%instruments 
detectors
performs non-demolition measurements resulting in a copy of sequence prepared by the Source can be approved only by specific properties of the sequence received. % Предположение о том, что один из приборов осуществляет неразрушающие измерения, результатом которых является копия последовательности, подготовленной источником, может подтверждаться только особыми свойствами принятой последовательности.
\end{block}

\end{frame}%FRAME 15

\subsection{Local Observables}
%FRAME 16
\begin{frame}{Qubit measurement}
The measurement of 3 observables % 3 наблюдаемых 
$\sigma_1$, $\sigma_2$, $\sigma_3$ is performed in three separate series. Density matrices of prepared states are same if Preparator's and Registrator's actions are coordinated. Differences of the readings of the counters % осуществляется  тремя отдельными сериями. Матрицы плотности приготовленных состояний одинаковы, если действия препаратора и регистратора согласованы. Разности показаний счетчиков 
$\frac{K_1-K_0}{K}$ give an unbiased estimate of the parameters of state% дают несмещенную оценку параметров состояния 
$\Delta p$, $d'$, $d'$, and
\[\hat{\rho}=p_0\roa{0}+p_1\roa{1}\]
If the measured values differ from zero more than by the value of the error depending on the length of the series, one can calculate the eigenvectors 
%Если измеренные значения отличаются от нуля больше, чем на величину зависящей от длины серии погрешности, можно  вычислить собственные векторы 
$\cat{0}$, $\cat{1}$ of density matrix that determine the pure states prepared by the Source. %матрицы плотности, определяющие приготавливаемые источником чистые состояния. 
  
  Preparator can exclude possibility of calculation of eigenvectors of density matrix by preparation of sequences with same numbers of states %Препаратор может  исключить возможность вычисления собственных векторов матрицы плотности, приготавливая последовательности с одинаковым количеством состояний 
  $\roa{0}$ and $\roa{1}$.  If in the sequence with length %  Если в последовательности длиной 
  $K$ difference between the numbers of states 0 and 1 is less than uncertainty % разность между числами состояний 0 и 1 меньше неопределенности 
  $\frac{1}{2\sqrt{K}}$, difference of the sequences of the results of measurements for each of observables from equidistributed ones is within the error, and it is not possible to calculate the space basis used by the Source. % отличие последовательностей результатов измерений каждой из наблюдаемых от равнораспределенной остается в пределах погрешности и вычислить используемый в источнике басис пространства состояний невозможно.

 \end{frame}%FRAME 16

\subsection{System Observables}

%FRAME 17
\begin{frame}{Measurement of a Qubit pair I}

\begin{block}{Local observables}
%Measurement

Let us choose three incompatible observables of an atom, let as an example those are the observable of the state number and two components of dipole moment represented by matrices % Выберем три несовместимых наблюдаемых атома, пусть это будут в качестве примера наблюдаемая номера состояния  и две компоненты дипольного момента, представляемые матрицами 
\[\hat{O}_A \in\left\{\sigma_3,\quad \sigma_1,\quad \sigma_2\right\}\]

In same way we choose three observables of the field with restriction of definition domain of those by means of projector % Точно так же выберем три наблюдаемые поля, ограничив их область определения с помощью проектора 
\(P_f=\eroa{\psi_0}+\eroa{\psi_\bot}
\)
to two-dimensional state space of the field mode entangled with the atom. 

Those can be observables represented by Pauli matrices: % на двумерное подпространство состояний моды поля, перепутанной с атомом. Это могут быть наблюдаемые, представляемые матрицами Паули:
\begin{equation}
	\begin{array}{l}
	\hat{s}_1=\ero{\psi_0}{\psi_\bot}+\ero{\psi_\bot}{\psi_0},\\ \hat{s}_2=-i\ero{\psi_0}{\psi_\bot}+i\ero{\psi_\bot}{\psi_0},\\ \hat{s}_3=\eroa{\psi_\bot}-\eroa{\psi_0}.
	\end{array}
\end{equation}
\end{block}
\begin{block}{System observables}
Then in arbitrary way we form three pairs of observables % образуем три пары наблюдаемых:
\[O_x \in\left\{\left[\sigma_x,s_x\right],\quad x=1,2,3\right\}\]
\end{block}
 \end{frame}%FRAME 17

\subsection{Set of  Series}

%FRAME 18
\begin{frame}{Measurement of Qubit pair II}
Three series of measurements for pairs of observables make it possible to obtain the values of three joint probability distributions % Три серии измерений для пар наблюдаемых  позволяют получить значения трех совместных распределений вероятностей
: \[\left\{\mathcal{R}_1,\mathcal{R}_2,\mathcal{R}_3\right\}\quad\mapsto\quad \left\{p^{\left\{x\right\}}_{b_a,b_f}, x=1,2,3;b_a,b_f=0,1\right\}\]
Those probability distributions provide calculation of reduced density matrices of the atom and the field:  %Эти распределения вероятностей обеспечивают вычисление редуцированных матриц плотности атома и поля:
\[\rho_a=\frac{1}{2}+\frac{1}{2}\sum_{x=1}^3\aver{\sigma_x}\sigma_x,\quad 
\rho_f=\frac{1}{2}+\frac{1}{2}\sum_{x=1}^3\aver{s_x}s_x.\]
By three-dimensional rotations %Трехмерными вращениями
\[\tilde{\sigma}_x=U^y_x\left(\aver{\sigma_x}\right){\sigma}_y,\quad \tilde{s}_x=U^y_x\left(\aver{s_x}\right){s}_y,\]
 corresponding to turn of orts   $\epsilon_3$ and $e_3$ in directions of the vectors $\aver{\sigma_x}$ and $\aver{s_x}$ local observables are separated to non-demolition % локальные наблюдаемые разделяются на неразрушающие 
 $\tilde{\sigma}_3$,  $\tilde{s}_3$ and over-classical $\tilde{\sigma}_{1,2}$,  $\tilde{s}_{1,2}$ ones.
  \end{frame}%FRAME 18

%FRAME 19
\begin{frame}{Measurement of Qubit pair III}

Two additional series of measurements of joint probability distributions for two cross pairs of over-classical observables %  Две дополнительных серии измерений совместных распределений вероятностей  двух перекрестных пар надклассических наблюдаемых
 $\tilde{\sigma}_{1}\tilde{s}_{2}$ and $\tilde{\sigma}_{2}\tilde{s}_{1}$ make it possible to calculate all the components of the matrix of covariance of over-classical observables % позволяют вычислить все компоненты матрицы ковариации надклассических наблюдаемых 

\[V_{a,b}=\aver{\tilde{\sigma}_{a}\tilde{s}_{b}}-\aver{\tilde{\sigma}_{a}}\aver{\tilde{s}_{b}}\]
Diagonalization of this matrix is accompanied by transformation of local bases to the form where Schmidt decomposition for the eigenvectors of density matrix of a composite state is canonical, and the system of equations that make it possible to determine all the parameters of density matrix of a composite state is reduced to the one considered above. %  Диагонализация этой матрицы сопровождается преобразованием локальных базисов к виду, в котором разложение Шмидта для собственных векторов матрицы плотности композитного состояния является каноническим, а система уравнений, позволяющих определить все параметры матрицы плотности композитного состояния, приводится к разобранной выше.

 Thus, complete reconstruction of a state of a composite system comprising a two-level atom and one mode of electromagnetic quantum field can be performed in result of measurement of 5 incompatible pairs of local observables. %  Таким образом, полное восстановление состояния композитной системы, состоящей из двухуровневого атома и одной моды электромагнитного квантового поля, может быть выполнено в результате измерения 5 несовместимых пар локальных наблюдаемых. 
 
 Complexity of Measurement
\[ 
	\log_2 \frac{5}{4}+\log_2 \frac{3}{4}-2\log_2 s \leq  C_4\left(s\right) \leq \log_2 \frac{5}{4}-2\log_2 s.
\]
Error that does not exceed  0.1 can be obtained in 5 series of measurements, from 19 to 25 events of measurement each. % Погрешность, не превышающая 0.1, может быть получена при проведении 5 серий измерений  длиной от 19 до 25 событий измерения каждая. 

\end{frame}%FRAME 19

\section{Conclusions}

%FRAME 20
\begin{frame}{Types of Measurement Tasks}
%\begin{itemize}
%	\item 
\begin{block}{}
Analysis of the process of accumulation of the results of measurements shows that the success of this process substantially depends on the possibility of coordination of actions of two participants of the process - Preparator who prepares the series of the states being measured and Registrator who chooses in each event of measurement one of incompatible observables. 

\end{block}

\begin{block}{}

% 	\item 
In the problems on information transfer or transformation, coordination of the source and the receiver makes it possible to get unambiguous reconstruction of states by one measurement (non-demolition measurements). 
 \end{block}

\begin{block}{}

%	\item 
 In the problems of research character or adjustment of equipment with coordination of the sequence of the prepared states and the sequence of the measured (incompatible) observables, even at use of entangled states, it is possible to decrease the needed number of the measurement events to the theoretical minimum. For a pair of qubits the minimal needed number is the measurement of 5 pairs of incompatible observables, for $N$ gubits -- $2^N+1$ observables. 
 
\end{block}

\begin{block}{}
 
%	\item 
The problems of information protection against interception differ by absence of coordination of actions of preparator and registrator, this substantially complicates the registrator's problem. It is shown that by an appropriate selection of coding sequences it is possible to achieve insolubility of the problem of reconstruction of the coding state set and as a result insolubility of the problem of intercept. 
\end{block}

%\end{itemize}

\end{frame}%FRAME 20

%FRAME 21
\begin{frame}{Complexity of Measurement}

\begin{block}{}
A new value, the complexity of measurement of a state, is used, this characterizes the number of the measurement events needed for solution of the reconstruction problem. 
 
\end{block}

\begin{block}{}
By the example of a pair of quibits formed by a two-level atom and a quantum mode coupled with the atom it is shown that measurement of 5 incompatible pairs of local observables is enough for reconstruction of an arbitrary state, and availability of prior information on the states that are used makes it possible to limit the measurement with 3 incompatible pairs of local observables. 
\end{block}

\end{frame}%FRAME 21

\begin{frame}{}%FRAME 22

{\huge 
\begin{center}
Thank You for attention

\end{center}}
\end{frame}%FRAME 22

\end{document}